\documentclass[usenatbib]{mn2e}
\usepackage{natbibmnfix,graphicx,times}

\newcommand{\bq}{\begin{equation}}
\newcommand{\eq}{\end{equation}}
\newcommand{\bqa}{\begin{eqnarray}}
\newcommand{\eqa}{\end{eqnarray}}

\newcommand{\kel}{\mbox{ K}}

\newcommand{\secinv}{\mbox{ s$^{-1}$}}

\newcommand{\hunits}{\mbox{ km s$^{-1}$ Mpc$^{-1}$}}

\newcommand{\lya}{Ly$\alpha$ }
\newcommand{\lyans}{Ly$\alpha$} 
\newcommand{\lyb}{Ly$\beta$ }
\newcommand{\lyn}{Ly$n$ }

\newcommand{\deriv}{{\rm d}}
\def\VEV#1{\left\langle #1\right\rangle} 

\newcommand{\olx}{\overline{x}}
\newcommand{\ulx}{\underline{x}}

\newcommand{\apj}{ApJ}
\newcommand{\apjl}{ApJ}

\newcommand{\aj}{AJ}
\newcommand{\mnras}{MNRAS}
\newcommand{\pasj}{PASJ}

\title[Lyman Series Photons in the IGM]{The scattering of Lyman-series photons in the intergalactic medium}

\author[S.~R. Furlanetto \& J.~R. Pritchard]{Steven R.  Furlanetto$^1$\thanks{Email: steven.furlanetto@yale.edu} and Jonathan R. Pritchard$^2$ \\
$^1$Yale Center for Astronomy and Astrophysics, Yale University, 260 Whitney Avenue, New Haven, CT 06520-8121 \\
$^2$California Institute of Technology, Mail Code 130-33, Pasadena, CA 91125, USA}

\begin{document}

\maketitle

\begin{abstract}
We re-examine scattering of photons near the \lya resonance in the intergalactic medium (IGM).  We first derive a general integral solution for the radiation field around resonance when spin diffusivity is ignored.  Our solution shows explicitly that recoil sources an absorption feature, whose magnitude increases with the relative importance of recoil compared to Doppler broadening.  This spectrum depends on the \lya line profile, but approximating it with the absorption profile appropriate to the Lorentzian wings of natural broadening accurately reproduces the results for a full Voigt profile so long as $T_K \la 1000 \kel$ in the IGM.  This approximation allows us to obtain simple analytic formulae for the total scattering rate of \lya photons and the accompanying energy exchange rate.  Our power series solutions converge rapidly for photons that redshift into the \lya resonance as well as for photons injected at line center.  We confirm previous calculations showing that heating through this mechanism is quite slow and probably negligible compared to other sources.  We then show that energy exchange during the scattering of higher-order Lyman-series photons can be much more important than naively predicted by recoil arguments.  However, the resulting heating is still completely negligible.
\end{abstract}
  

\section{Introduction} \label{intro}

The radiative transfer of photons near the \lya resonance is crucial to understanding the high-redshift intergalactic medium (IGM), both because it determines the spin temperature of the 21 cm transition \citep{wouthuysen52, field58} and because it affects the thermal history \citep{madau97, chen04}.  

The radiation field near this resonance has been examined a number of times in recent years.  The earliest treatments ignored radiative transfer and assumed that the spectrum was featureless around the line.  \citet{chen04} were the first to solve (numerically) an approximate form of the radiative transfer equation in this context (following \citealt{basko81} and \citealt{rybicki94}).  They showed that, if photons redshift toward the resonance, the spectrum develops an asymmetric absorption feature.  As we will see explicitly below, the absorption feature is sourced by recoil in the scattering process:  each scattering deposits an average energy $\Delta E = (h \nu_\alpha)^2/(m_p c^2)$, where $\nu_\alpha$ is the rest frequency of the \lya line.  Thus photons lose energy faster near the center of resonance, where they scatter more.  To compensate for this increased ``flow" speed, continuity requires that the amplitude of the background must decrease near resonance.  This affects the scattering rate of \lya photons and hence the spin temperature of the IGM.  \citet{hirata05} expanded on this method by showing how to account for the hyperfine structure of the \lya line (see below).

An alternative to the numerical approach of \citet{chen04} and \citet{hirata05} is to approximate the spectrum analytically.  This has a long history in resonant radiative transfer; \citet{hummer92} summarize many of the advances.  Of particular interest to our problem is the treatment of \citet{grachev89}, who derived an analytic solution for the spectrum around a resonant transition when recoil is included.  The analytic solution was obtained by approximating the absorption profile using the form appropriate for scattering in the Lorentzian wings provided by natural broadening.  This assumption is valid when the optical depth is extremely large and the Doppler broadening relatively small.  Most recently, \citet{chuzhoy06} rediscovered this solution and applied it to the problem of \lya transfer in the high-redshift IGM.  In \S \ref{radfield}, we will show how these two kinds of solutions relate and evaluate when the analytic approximation is valid.  We also compute the radiation field in such a way that the role of recoil becomes obvious.  We examine the resulting total scattering rate in \S \ref{scatt} and show that the approximate form proposed by \citet{chuzhoy06} is a reasonably good match to the full numeric result.

The line shape is also crucial for estimating the rate at which energy is transferred between the gas and the photon field.  As described above, recoil during each scattering deposits some energy in the gas.  If this were the sole mechanism for energy exchange, the IGM would rapidly be heated above the cosmic microwave background (CMB) temperature \citep{madau97}.  However, the absorption feature actually cancels almost all of this heating.  Consider a photon on the blue side of the line.  This will be preferentially scattered by an atom moving \emph{away} from the photon (so that it appears closer to resonance).  The atom will then re-emit the photon isotropically in its frame; in the IGM frame, the photon will therefore lose an energy $\sim h \Delta \nu_D$, where $\Delta \nu_D$ is the Doppler width of the transition.  Photons that scatter on the red side, on the other hand, will tend to gain energy.  The absorption feature develops so that this scattering ``diffusivity" compensates for the recoil (i.e., so that more scattering occurs redward than blueward of the \lya transition).  The net energy transfer is therefore much slower than naively expected \citep{chen04, rybicki06, meiksin06}.  

By employing their analytic approximation to the radiation field, \citet{chuzhoy06-heat} took a step toward finding a simple solution for the net heating rate.  In \S \ref{lyaheat}, we take their approach further by deriving a fully analytic solution for heating by photons redshifting into the \lya resonance as well as an approximate solution for photons injected at line center  (either through recombinations or cascades from higher \lyn transitions).  This allows us to examine how the heating rate varies with IGM temperature and optical depth.    

Of course, photons can redshift into any of the \lyn resonances in the IGM.  After a few scatterings, these photons are destroyed through cascades to lower levels \citep{hirata05, pritchard05}.  The scattering rate is so small that recoil heating is negligible; however, all of the scatterings occur on the blue side of the line, so each deposits some fraction of the atom's thermal energy in the gas as well.  \citet{chuzhoy06-heat} examined the analogous process in deuterium and found that it can be relatively strong.  In \S \ref{lyn}, we show that the heating rate for \lyn photons is tiny even when frequency drift is included, because the photons scatter so far in the blue wing of the line.

\citet{hirata05} and \citet{chuzhoy06} examined how spin exchange affects the radiation spectrum.  Because \lya transitions modify the ground-state hyperfine level populations \citep{wouthuysen52, field58}, the photons can also increase or decrease their frequency during each scattering by an amount corresponding to the energy defect of the 21 cm transition.  This affects the flow rate of photons through the resonance (or more precisely the diffusivity) and hence the spectrum.  However, because the level populations themselves depend on the \lya scattering rate, and because the mean energy exchange per scattering depends on the level populations, including this effect on the spectrum requires an iterative solution.  Fortunately, it is a small effect except at extremely low temperatures.  Because we are most interested in the gross behavior of the scattering and heating rates, we will neglect the spin diffusivity.  When required, including it is relatively easy; the steps are outlined in \citet{furl06-review} (see also \citealt{hirata05}).

In our numerical calculations, we assume a cosmology with $\Omega_m=0.26$, $\Omega_\Lambda=0.74$, $\Omega_b=0.044$, and $H=100 h \hunits$ (with $h=0.74$), consistent with the most recent measurements \citep{spergel06}.

\section{The Radiation Field Near the \lya Resonance}
\label{radfield}

We let $J$ be the comoving angle-averaged specific intensity (in units of photons per area per steradian).  The equation of radiative transfer is (neglecting atomic recoil for the moment)
\bqa
\frac{1}{c n_H \chi_\alpha} \, \frac{\partial J}{\partial t} & = & -\phi(\nu) \, J + H \nu_\alpha \, \frac{\partial J}{\partial \nu} + \int \deriv \nu' \, R(\nu,\nu') \, J(\nu') \nonumber \\
& & + C(t) \psi(\nu),
\label{eq:rt}
\eqa
where $n_H$ is the hydrogen density, $\sigma_{\alpha}(\nu)=\chi_{\alpha} \phi(\nu)$ is the absorption cross section, $\chi_{\alpha} = (\pi e^2/m_e c) f_{\alpha}$, $f_{\alpha}$ is the absorption oscillator strength, and $\phi(\nu)$ is the line profile.  For our purposes, $\phi$ is given by the Voigt profile (which includes both collisional and natural broadening),
\begin{equation}
\phi(x) = \frac{a}{\pi^{3/2}} \int_{-\infty}^{\infty} \deriv t \, \frac{e^{-t^2}}{a^2 + (x-t)^2},
\label{eq:voigt}
\end{equation}
with $a=\Gamma/(4 \pi \Delta \nu_D)$, $\Gamma$ the inverse lifetime of the upper state, $\Delta \nu_D/\nu_0=(2 k_B T_K/m c^2)^{1/2}$ the Doppler parameter, $T_K$ the gas temperature, and $x \equiv (\nu-\nu_0)/\Delta \nu_D$ the normalized frequency shift.  The first term on the right-hand side of equation~(\ref{eq:rt}) describes absorption, the second the Hubble flow, and the third re-emission following absorption.  The redistribution function $R(\nu,\nu')$ gives the probability that a photon absorbed at frequency $\nu'$ is re-emitted at frequency $\nu$.  The approximate form $R_{\rm II}(\nu,\nu')$ \citep{henyey41, hummer62}, which assumes a Voigt profile with coherent scattering in the rest frame of the absorbing atom, is often used (see \S \ref{lyn}).  We must, however, also include recoil \citep{basko81} and, for exact calculations, spin exchange \citep{hirata05, chuzhoy06}.  The last term describes injection of new photons:  $C$ is the rate at which they are produced and $\psi(\nu)$ is their frequency distribution.  

This integro-differential equation simplifies considerably if we assume that the background spectrum is smooth on the scale of the average frequency change per scattering (which is $\Delta x < 1$; see \S \ref{lyn}).  If we neglect spin exchange in this Fokker-Planck approximation, equation~(\ref{eq:rt}) becomes \citep{rybicki94}
\begin{equation}
\frac{\deriv}{\deriv x} \left\{ \phi(x) \frac{\deriv J}{\deriv x} + 2 [\eta \phi(x) + \gamma] J(x) \right\} + C \psi(x) = 0,
\label{eq:rt-fp}
\end{equation}
where the Sobolev parameter $\gamma=\tau_{\rm GP}^{-1}$, $\tau_{\rm GP}$ is the total \citet{gunn65} optical depth of the \lya transition, and we have included atomic recoil through the parameter $\eta = (h \nu_\alpha^2)/(m c^2 \Delta \nu_D)$ \citep{basko81}; this is the mean (normalized) frequency drift per scattering from recoil.  Unfortunately, equation~(\ref{eq:rt-fp}) is not uniquely specified because there is some freedom in the drift and diffusivity imposed in the Fokker-Planck method.  The form above matches that of \citet{rybicki94}; however, it does not obey detailed balance, which requires $\eta \rightarrow \eta - 1/(x + x_\alpha) \approx \eta - 1/x_\alpha$ where $x_\alpha \equiv \nu_\alpha/\Delta \nu_D$ \citep{rybicki06}.  The correction is unimportant when $k_B T_K \ll h \nu_\alpha$ but is easily included in the analysis.  Other forms of the Fokker-Planck approximation have been examined by \citet{meiksin06}.  It is also straightforward to include the drift and diffusivity sourced by hyperfine mixing in this formalism, so long as the spin temperature is known \citep{hirata05, chuzhoy06}.

It is useful now to pause and note explicitly the scalings of the basic parameters of this problem; they will become useful later.  We have $\Delta \nu_D \propto T_K^{1/2}$, so $a \propto T_K^{-1/2}$ and $\eta \propto T_K^{-1/2}$.  The Sobolev parameter $\gamma \propto (1+z)^{-3/2}$ in the high-redshift limit.

We will consider two sets of boundary conditions for equation~(\ref{eq:rt-fp}).  First, we let photons redshift into the resonance from large frequencies, with no injection term.  To describe this we let $J_\infty > 0$ be the specific intensity as $x \rightarrow \infty$ and set $C=0$.  The second case allows injection at line center, so $C \psi(x) = C \delta(x)$,\footnote{Even if the initial Lyman-series absorption occurs well blueward of line center, the \lya photon that results from the cascade will be injected near line center because the atom passes through several intermediate states, each of which has a small natural width.} and sets $J_\infty=0$.  In this case, we define $J_{-\infty}$ to be the average intensity as $x \rightarrow -\infty$ as well.  In either scenario, equation~(\ref{eq:rt-fp}) is easy to integrate once, leaving us with a first order ordinary differential equation.

\begin{figure}
\begin{center}
\resizebox{8cm}{!}{\includegraphics{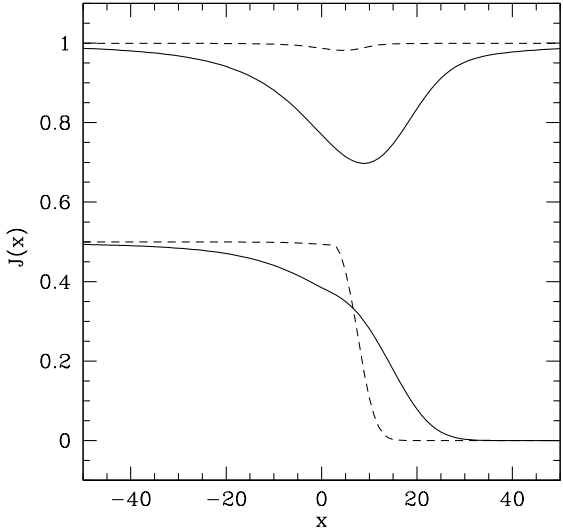}}\\%
\end{center}
\caption{Background radiation field near the \lya resonance at $z=10$, assuming a Voigt line profile.  The upper and lower sets are for photons redshifting from infinity and photons injected at line center, respectively.  (The former are normalized to $J_\infty$; the latter have $J_{-\infty}=1/2$.)  The solid and dashed curves take $T_K=10$ and $1000 \kel$, respectively.}
\label{fig:jnu}
\end{figure}

The formal solution is most transparently obtained by changing variables to \citep{hummer92}
\begin{equation}
\sigma(x) = \int_0^x \frac{\deriv x'}{\phi(x')},
\label{eq:sigdefn}
\end{equation}
so that equation~(\ref{eq:rt-fp}) becomes
\begin{equation}
\frac{\deriv J}{\deriv \sigma} + 2 (\eta \phi + \gamma) J = 2 K,
\label{eq:rt-sig}
\end{equation}
where $K = \gamma J_\infty$ for the continuous case, $K=C$ for injected photons if $x<0$, and $K=0$ for injected photons with $x>0$.  Obviously
\begin{equation}
\exp \left[ 2 \eta \int_0^{\sigma} \phi(\sigma') \deriv \sigma' + 2 \gamma \sigma \right]
\label{eq:intfac}
\end{equation}
is an integrating factor for this equation, from which the solution follows immediately.  For injected photons with $x>0$ (so that $K=0$), it has the simple form
\begin{equation}
J(x) = J(0) \exp \left[ - 2 \eta x - 2 \gamma \int_0^x \frac{\deriv x'}{\phi(x')} \right],
\label{eq:soln-inj}
\end{equation}
where $J(0)$ is determined by continuity.  

\begin{figure}
\begin{center}
\resizebox{8cm}{!}{\includegraphics{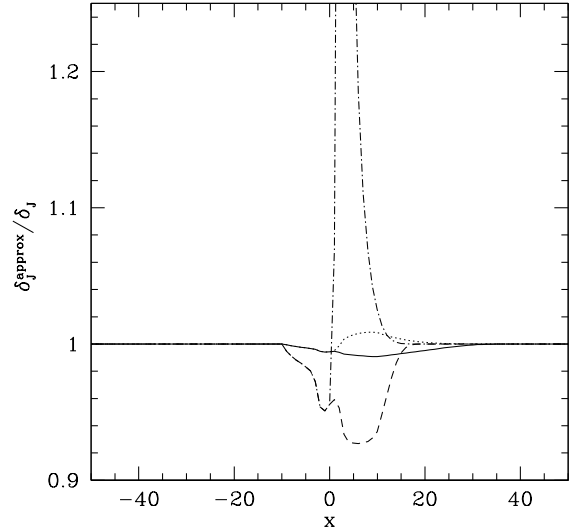}}\\%
\end{center}
\caption{Ratio of $\delta_J$ in the ``wing" approximation to the exact results (using a Voigt profile).  The solid and dashed curves assume a flat background spectrum and take $T_K=10$ and $1000 \kel$, respectively.  The dotted and dot-dashed curves assume injection at line center, with $T_K=10$ and $1000 \kel$, respectively.  In this case, we set $\delta_J=J/J_{-\infty}$ for $x>0$.  All curves assume $z=10$.}
\label{fig:jnurat}
\end{figure}

A formal solution can also be written for $K>0$, but in this case an alternate form is more physically illuminating.  Here it is the absorption spike that is most interesting.  To isolate its properties, we define $\delta_J \equiv (J_\infty - J)/J_\infty$;\footnote{For injected photons, $J_\infty=0$, of course; then we make the substitution $J_\infty \rightarrow J_{-\infty}$ in the definition.  We will see that $J_\infty = J_{-\infty}$ for a redshifting continuum.}  note then that $\delta_J > 0$.  The transfer equation takes the form
\begin{equation}
\phi \frac{\deriv \delta_J}{\deriv x} + 2 (\eta \phi + \gamma) \delta_J = 2 \eta \phi.
\label{eq:rt-dj}
\end{equation}
This has the same structure as the previous version, except that the sourcing term on the right-hand side depends on $x$.  The same integrating factor yields the solution
\begin{equation}
\delta_J(x) = 2 \eta \int_0^\infty \deriv y \exp \left[ - 2 \eta y - 2 \gamma \int_{x-y}^{x} \frac{\deriv x'}{\phi(x')} \right].
\label{eq:dj-soln}
\end{equation}
This form makes it obvious that recoil sources the absorption spike.  If the scattering was purely coherent, the gas and radiation field could not transfer any radiation and the spectrum would remain flat (see, e.g., \citealt{hummer92}).  By sapping energy from each scattered photon, recoil increases the rate at which they redshift across the resonance.  This increase in the ``flow velocity" must be balanced by a corresponding decrease in the photon flux near the resonance.  

We show some example spectra in Figure~\ref{fig:jnu}, assuming that $\phi(x)$ has a Voigt profile (see also \citealt{chen04}).  The upper curves assume that photons redshift into resonance from infinity; as expected, an absorption feature develops.  It deepens at small temperatures, because, in that case, the energy lost from recoil is large compared to the energy lost in each scattering (or $\eta$ is relatively large).  The lower curves assume injection at line center.  In this case, the spectrum spreads to large positive $x$ when $T_K$ decreases.

This numerical solution is, of course, identical to those presented by \citet{chen04} and \citet{hirata05}, once the appropriate line profiles, drifts, and diffusivities are inserted.  It is also a more general form of the solutions provided by \citet{hummer92} (who neglected the recoil term) and \citet{chuzhoy06}.  The latter made the approximation (following \citealt{chugai80, chugai87, grachev89}) that $\phi(x) \approx a/(\pi x^2)$, which is only accurate at $|x| \gg 1$.  We will refer to this as the ``wing" approximation for convenience.  This approximation allows the integrals over $\phi^{-1}$ to be performed analytically \citep{grachev89, chuzhoy06}.  For injected photons with $x>0$, it is
\bq
J(x) = J(0) \exp \left( - 2 \eta x - \frac{2 \pi}{3} \, \frac{\gamma x^3}{a} \right), \label{eq:jinj-wing}
\eq
while for a flat background or injected photons with $x<0$, 
\bq
\delta_J(x) = 2 \eta \int_0^\infty \deriv y \, \exp \left[ - \frac{2 \pi \gamma}{3 a} (y^3 - 3 y^2 x + 3 y x^2) - 2 \eta y \right].
\label{eq:dj-wing}
\eq
This explains the discrepancy between the existing numeric and analytic results:  the latter do not apply near the Doppler core of the profile.  

\begin{figure}
\begin{center}
\resizebox{8cm}{!}{\includegraphics{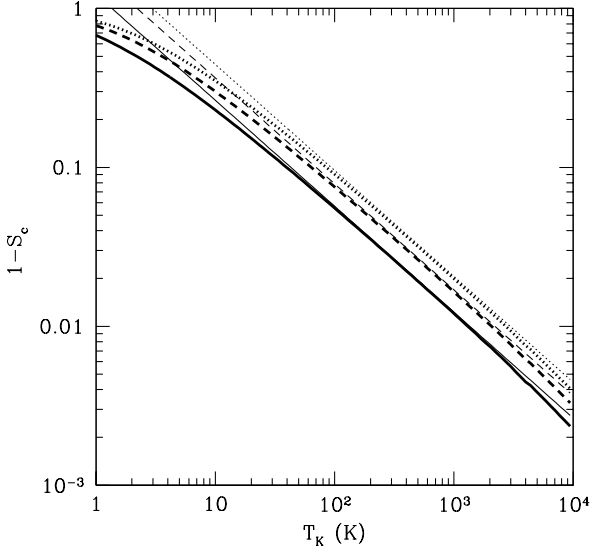}}\\%
\end{center}
\caption{Scattering integral as a function of IGM temperature.  The thick solid, dashed, and dotted curves show $(1-S_c)$ for a Voigt profile at $z=10,\,20$, and $30$.  The thin curves show the corresponding quantities using only the first-order term in equation~(\ref{eq:s-wing-power}).  }
\label{fig:sctk}
\end{figure}

Figure~\ref{fig:jnurat} shows the ratio of the approximate analytic form of \citet{grachev89} and \citet{chuzhoy06} to the exact spectra (computed with a Voigt profile).  When the temperature is small (solid and dotted curves), the approximation is an excellent one.  However, it begins to break down at large temperatures:  for example, in the continuous case, it underpredicts $\delta_J$ by $\sim 10\%$ at the center of the absorption spike when $T_K=1000 \kel$.  This is because the effective natural width decreases with temperature, so the thermal broadening becomes relatively more important in higher-temperature gas.  For injected photons, the wing approximation slightly shifts the curves to the right; when the decline at $x>0$ is sharp (as in the warm gas), the fractional deviation can be large.  However, we find that the wing approximation is generally an excellent one.  In the next two sections, we will use this analytic form to study the scattering and heating rates, extending the approach of \citet{chuzhoy06, chuzhoy06-heat}.

\section{The \lya Scattering Rate}
\label{scatt}

The total rate at which \lya photons scatter (per hydrogen atom) is
\begin{equation}
P_\alpha = 4 \pi \chi_\alpha \int_{-\infty}^\infty \deriv \nu \, J(\nu) \phi(\nu),
\label{eq:palpha}
\end{equation}
where $J$ is now in proper units.  Because each scattering can exchange hyperfine states, this rate is crucial for determining the spin temperature of the 21 cm transition in the IGM \citep{wouthuysen52, field58, madau97, chen04, hirata05, chuzhoy06}.  The Wouthuysen-Field coupling strength can be written as (e.g., \citealt{furl06-review})\footnote{For injected photons, one must substitute $J_\infty \rightarrow J_{-\infty}$.}
\begin{equation}
x_\alpha = \frac{16 \pi \chi_\alpha J_\infty}{27 A_{10}} \, \frac{T_\star}{T_\gamma} \, S_\alpha,
\label{eq:xalpha}
\end{equation}
where $A_{10}=2.85 \times 10^{-15} \secinv$ is the spontaneous emission coefficient of the 21 cm transition, $T_\star=0.068 \kel$ is the energy defect of that transition, $T_\gamma$ is the CMB temperature, and 
\begin{equation}
S_\alpha \equiv \int_{-\infty}^\infty  \deriv x \, \phi(x) \frac{J}{J_\infty}
\label{eq:salpha}
\end{equation}
depends only on the \emph{shape} of the background spectrum.  Note that $S_\alpha < 1$, because recoil always induces an absorption feature.

\begin{figure}
\begin{center}
\resizebox{8cm}{!}{\includegraphics{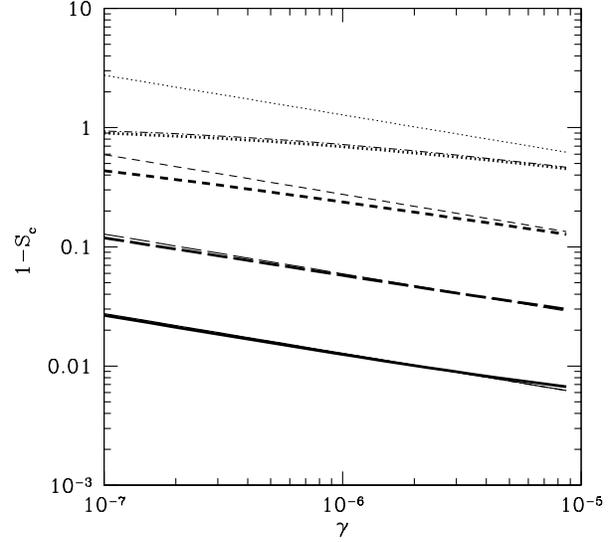}}\\%
\end{center}
\caption{Scattering integral as a function of $\gamma=\tau_{\rm GP}^{-1}$.  The thick curves show $(1-S_c)$ computed numerically for a Voigt profile, while the thin curves show the corresponding quantities using only the first order term in equation~(\ref{eq:s-wing-power}).  The dotted, short-dashed, long-dashed, and solid curves take $T_K=1,\,10,\,10^2$, and $10^3 \kel$, respectively.  The thin dot-dashed curve shows the approximate form proposed by \citet{chuzhoy06} for $T_K=1 \kel$.  }
\label{fig:scgam}
\end{figure}

In general, $S_\alpha$ must be computed numerically; even in the wing approximation, there is no closed-form analytic solution.  However, recall that $\phi(x)$ is sharply peaked around $x=0$, while $J$ varies slowly near resonance (even in the injected case).  Thus we can approximate $J \approx J(0)$ everywhere inside the integral; from the normalization of $\phi$ we thus have
\begin{equation}
1 - S_\alpha \approx \delta_J(0).
\label{eq:dsalpha}
\end{equation}
In the wing approximation, this is easily computed from equation~(\ref{eq:dj-wing}): 
\begin{eqnarray}
1 - S_\alpha & \approx &\frac{4 \alpha}{9} \left[ 3^{2/3} \pi {\rm Bi}\left( -\frac{2 \alpha}{3^{1/3}} \right) \right.\nonumber \\ 
& & \left. + (3 \alpha^2) \,  _1F_2 \left( 1; \frac{4}{3}, \frac{5}{3} ; - \frac{8 \alpha^3}{27} \right)  \right],
\label{eq:s-wing} \\
 & \approx & \frac{4 \pi}{3 \sqrt{3} \Gamma(2/3)} \alpha - \frac{8 \pi}{3 \sqrt{3} \Gamma(1/3)} \alpha^2 + \frac{4}{3} \alpha^3 + ...,
 \label{eq:s-wing-power}
\end{eqnarray}
where ${\rm Bi}(x)$ is an Airy function, $_1 F_2$ is a hypergeometric function, and
\begin{equation}
\alpha = \eta \left( \frac{3 a}{2 \pi \gamma} \right)^{1/3} \approx 0.717 T_K^{-2/3} \left( \frac{10^{-6}}{\gamma} \right)^{1/3},
\label{eq:alpha}
\end{equation}
where $T_K$ is in degrees Kelvin.  When $\alpha$ is small, we therefore have $(1 - S_\alpha) \propto T_K^{-2/3} \tau_{\rm GP}^{1/3}$.  This scaling gives some intuition for how the coupling strength varies in the IGM.  As in Figure~\ref{fig:jnu}, the absorption spike becomes less and less significant as $T_K$ increases; thus we must have $S_\alpha \rightarrow 1$ (its value without recoil) in a warm IGM.  The perturbation increases with optical depth because that increases the number of scatterings (and hence the energy loss due to recoil).

We show the dependence of $(1-S_\alpha)$ on temperature in Figure~\ref{fig:sctk} and the dependence on the Sobolev parameter (or optical depth) in Figure~\ref{fig:scgam}.  The thick curves show the numeric solution for a Voigt line profile and for a continuous background spectrum, which we denote $S_c$.  The case with photons injected at the line center has a nearly identical scattering integral, because $\delta_J(0)$ is identical in the two cases; only at high temperatures does the structure around resonance matter.  The thin curves show the first-order (in $\alpha$) approximation of equation~(\ref{eq:s-wing-power}).  We see that this provides an excellent match at $T_K \ga 10 \kel$, especially when $\gamma$ is relatively large (i.e., at lower redshifts).  

Note that we have actually made three approximations here:  (i) a constant $J$ across the line; (ii) the wing approximation; and (iii) the small $\alpha$ approximation.  The culprit at small $T_K$ is the third.  Here $\alpha$ is large and the power series approximation breaks down.  However, even including just terms up to $\alpha^3$ dramatically improves the estimate, with errors $\la 10\%$ so long as $T_K > 2 \kel$.  This demonstrates that the first approximation is an excellent one here:  at such small temperatures, $\phi(x)$ is extremely sharply peaked.  The second is equally good.  \citet{chuzhoy06} proposed the fit
\begin{equation}
\delta_J(0) \approx 1 - \exp (-1.12 \alpha),
\label{eq:csfit}
\end{equation}
which retains the first order behavior of $\delta_J(0)$ at small $\alpha$ (and hence is reasonably accurate) and fits the behavior for $\alpha \sim 1$ much better.  The thin dot-dashed curve in Figure~\ref{fig:scgam} shows how well this approximation does at $T_K=1 \kel$; it typically differs from the exact solution by $\sim 5\%$.

The overall agreement worsens at large temperatures as well.  Here $\alpha$ is small, so the power series in equation~(\ref{eq:s-wing-power}) converges rapidly and approximation (iii) is excellent.  The problem lies instead with the other two.  As we have seen, the wing approximation breaks down once $T_K$ exceeds $\sim 1000 \kel$.  This causes up to a $10\%$ underestimate of $\delta_J(0)$.  At the same time, approximation (i) breaks down and the region around resonance starts to contribute to the scattering integral.  This causes a $\la 20\%$ overestimate of $(1-S_c)$ compared to the exact result; fortunately, these two effects partially cancel.  

In summary, the fit proposed by \citet{chuzhoy06} (in eq.~\ref{eq:csfit}) is an excellent approximation to $S_\alpha$ unless high accuracy is required.  However, we emphasize that, in order to include spin transfer properly, one must still use an iterative procedure \citep{hirata05, chuzhoy06}.  

\begin{figure*}[!t]
\begin{center}
\resizebox{8cm}{!}{\includegraphics{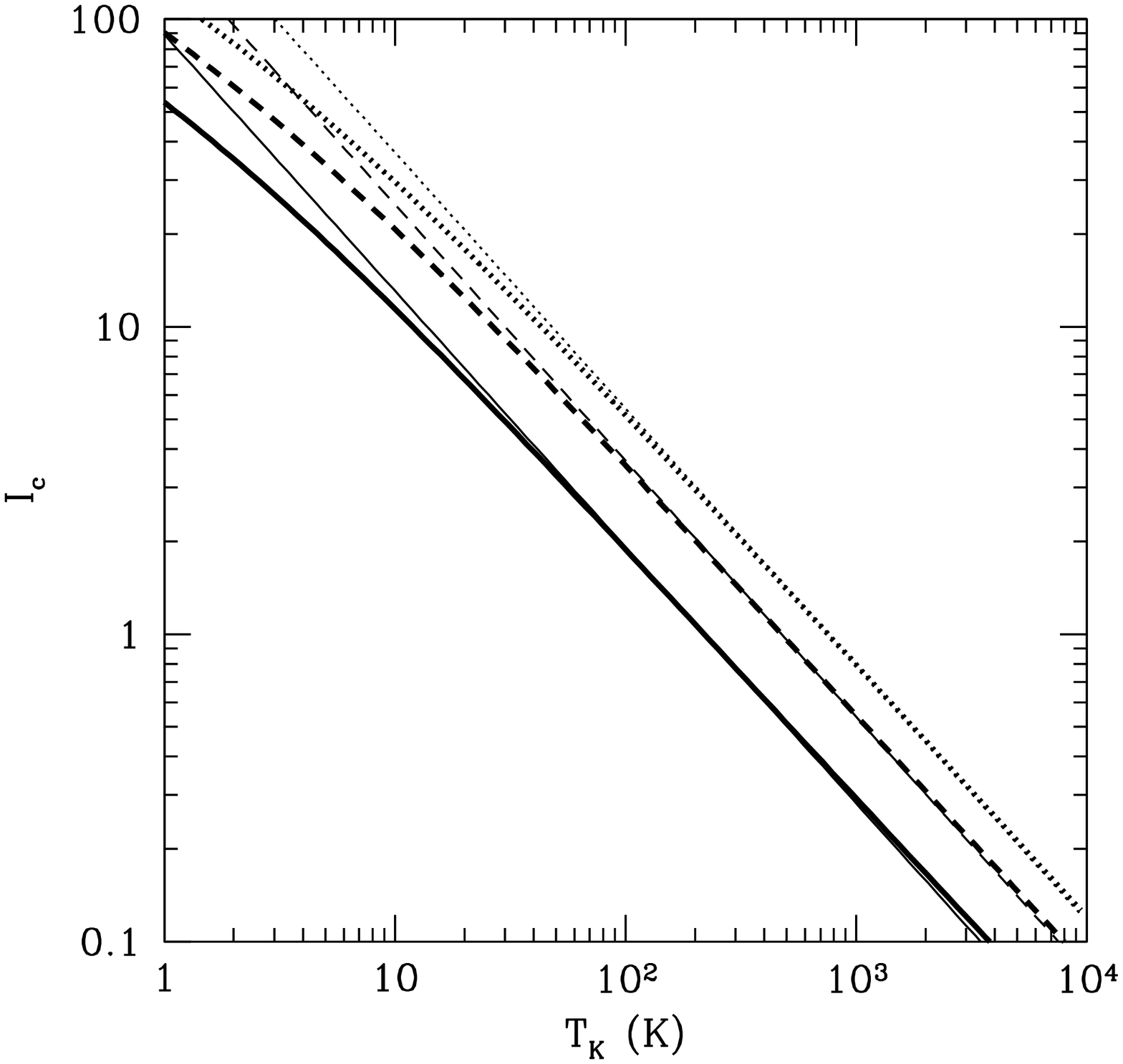}}
\hspace{0.13cm}
\resizebox{8cm}{!}{\includegraphics{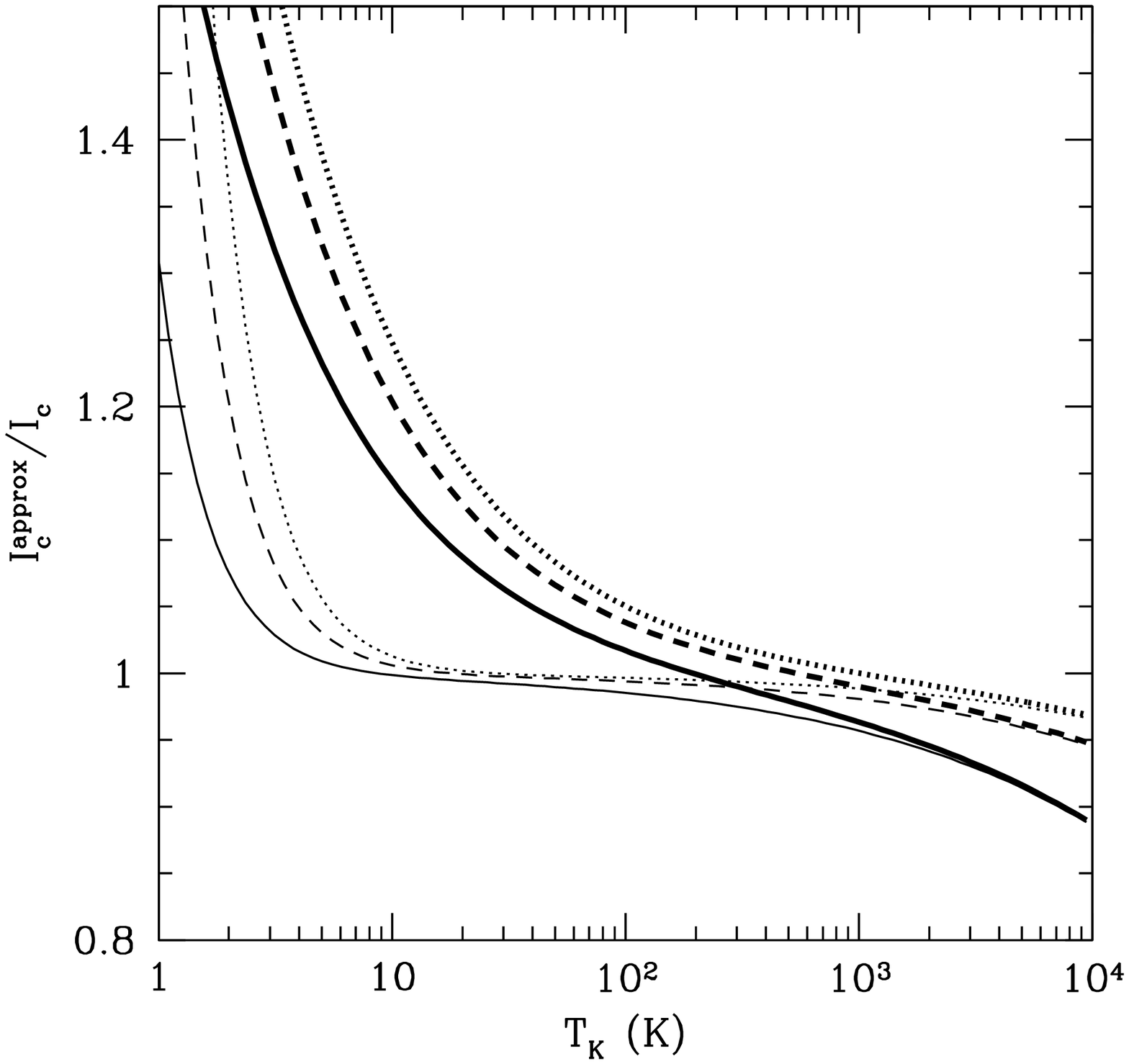}}
\end{center}
\caption{Heating integral for continuous injection.  \emph{Left panel:}  The thick solid, dashed, and dotted curves show $I_c$ for a Voigt profile at $z=10,\,20$, and $30$.  The thin curves show the corresponding quantities using only the first order term in equation~(\ref{eq:icapprox}).  \emph{Right panel:}  Ratio of the power series approximation to $I_c$ (using the wing approximation) to the exact value.  The thick and thin curves retain terms to order $\beta$ and $\beta^3$, respectively.}
\label{fig:ictk}
\end{figure*}

\section{Heat Exchange From \lya Scattering}
\label{lyaheat}

\subsection{Continuous Background} \label{cont-heat}

The rate at which the radiation field deposits energy in the gas (per unit volume) is \citep{chen04}
\begin{equation}
\Gamma = \frac{4 \pi H h \nu_\alpha}{c} \int_{-\infty}^{\infty} \deriv \nu \, (J_\infty - J),
\label{eq:heatrate}
\end{equation}
where we have assumed $\nu \approx \nu_\alpha$ across the absorption feature and $J$ is again in proper units.  The physical interpretation of this form is straightforward:  in the absence of scattering, the absorption feature would redshift away to infinity.  To keep it in place, the photons must lose energy at the rate given in equation~(\ref{eq:heatrate}).  More formally, it can be derived from the average energy exchange per scattering \citep{chuzhoy06-heat} through integration by parts and the use of equation~(\ref{eq:rt-fp}).

Thus the heating rate depends on
\begin{eqnarray}
I_c & = & \int_{-\infty}^{\infty} \deriv x \, \delta_J(x), \label{eq:icdefn} \\
& = & 2 \eta \int_0^\infty \deriv y \, e^{-2 \eta y} \int_{-\infty}^\infty \deriv x \exp \left[ -2 \gamma \int_{x-y}^x \frac{\deriv x'}{\phi(x')} \right].
\label{eq:icsub}
\end{eqnarray}
This cannot be done in closed form for an arbitrary line profile, but the accuracy of the wing approximation makes it extremely useful in understanding the solution.\footnote{Note that \citet{chuzhoy06-heat} apparently calculated the heating rate numerically in the wing approximation.}  In this case, both integrals can be done analytically, yielding
\begin{equation}
I_c = \left( \frac{4}{\pi} \right)^{-1/6} \pi^{3/2} \left( \frac{a}{\gamma} \right)^{1/3} \beta \left[ {\rm Ai}^2(-\beta) + {\rm Bi}^2(-\beta) \right],
\label{eq:ic-wing2}
\end{equation}
where
\begin{equation}
\beta = \eta \left( \frac{4a}{\pi \gamma} \right)^{1/3} =  0.99 T_K^{-2/3} \left( \frac{\gamma^{-1}}{10^6} \right)^{1/3},
\label{eq:betadefn}
\end{equation}
and ${\rm Ai}(x)$ and ${\rm Bi}(x)$ are the Airy functions and $T_K$ is in degrees Kelvin.  Note that this solution is \emph{exact} within the wing approximation.

We can again find a simple and useful approximation by expanding in powers of $\beta$.  We find
\bqa
I_c & \approx & 3^{1/3} \left( \frac{2 \pi}{3} \right)^{5/3} \left( \frac{a}{\gamma} \right)^{1/3} \left[ \frac{\beta}{\Gamma^2(2/3)} \beta - \frac{3^{1/3} \beta^2}{\Gamma(1/3) \Gamma(2/3)} \right. \nonumber \\ 
& & \left. + \frac{3^{2/3} \beta^3}{\Gamma^2(1/3)} + ... \right].
\label{eq:icapprox}
\eqa
Thus, we see $I_c \propto T_K^{-5/6} \gamma^{-2/3}$; because $\gamma=\tau_{\rm GP}^{-1} \propto (1+z)^{-3/2}$ at high redshifts, we expect $I_c \propto (1+z)$ at fixed temperature.  The heat input per atom per Hubble time (at constant $J_\infty$) is therefore $\Delta T \propto H \Delta \nu_D I_c/(n_{\rm HI} H) \propto T_K^{-1/3} (1+z)^{-2}$.  These scalings are close to those estimated by \citet{chuzhoy06-heat} from their numerical results.

\begin{figure}
\begin{center}
\resizebox{8cm}{!}{\includegraphics{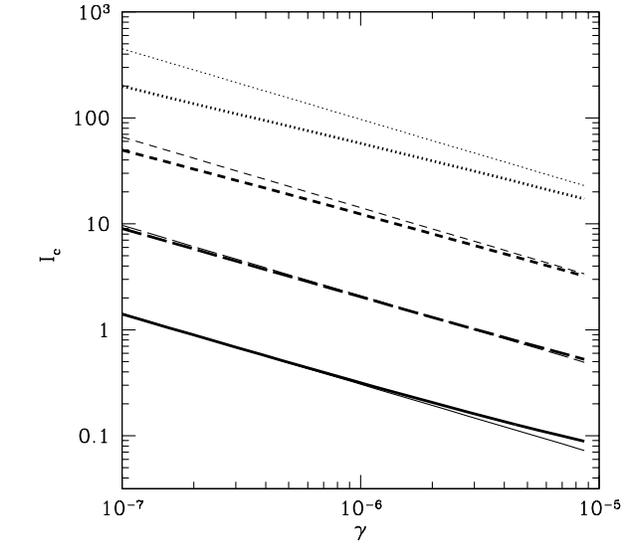}}\\%
\end{center}
\caption{Heating integral for continuous injection.  The thick curves show $I_c$ computed numerically for a Voigt profile, while the thin curves show the corresponding quantities using only the first order term in equation~(\ref{eq:icapprox}).  The dotted, short-dashed, long-dashed, and solid curves take $T_K=1,\,10,\,10^2$, and $10^3 \kel$, respectively.}
\label{fig:icgam}
\end{figure}

We show our solution for $I_c$ as a function of $T_K$ in the left panel of Figure~\ref{fig:ictk} and as a function of $\gamma$ in Figure~\ref{fig:icgam}.  In each of these panels, the thick curves use the full Voigt profile, while the thin curves use the first-order term (in $\beta$) of equation~(\ref{eq:icapprox}); we expect the latter to be valid when $T_K \ga 10 \kel$.  The right panel of Figure~\ref{fig:ictk} shows the ratio of the approximate and exact  solutions; here the thick curves retain only the lowest order term, while the thin curves include terms up to $\beta^3$:  these are necessary for $T_K \la 10 \kel$.  As before, the expansion in equation~(\ref{eq:icapprox}) converges rapidly at higher temperatures, but the wing approximation begins to break down.

Obviously the predicted scalings are reasonably accurate; the heating rate decreases with temperature (because recoil is relatively inefficient) and increases with $\tau_{\rm GP}$ (along with the scattering rate).  The higher-order terms, and the Voigt profile, slightly decrease the dependence on these parameters.  As shown by \citet{chen04}, \lya heating is probably slow compared to other processes, and the wing approximation (in the full analytic expression for small temperatures and the power series form otherwise) should be adequate for most purposes.

\begin{figure*}[!t]
\begin{center}
\resizebox{8cm}{!}{\includegraphics{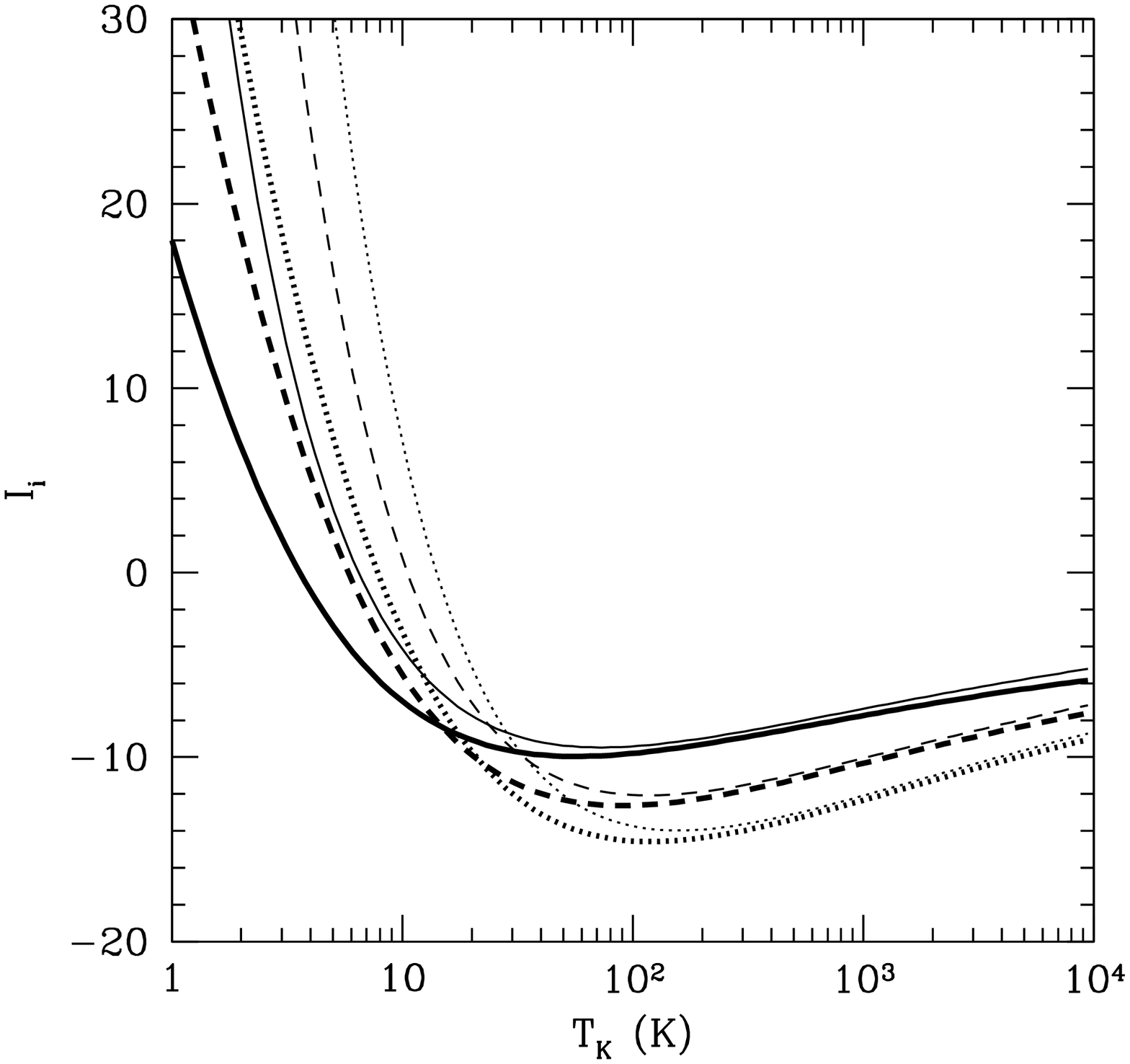}}
\hspace{0.13cm}
\resizebox{8cm}{!}{\includegraphics{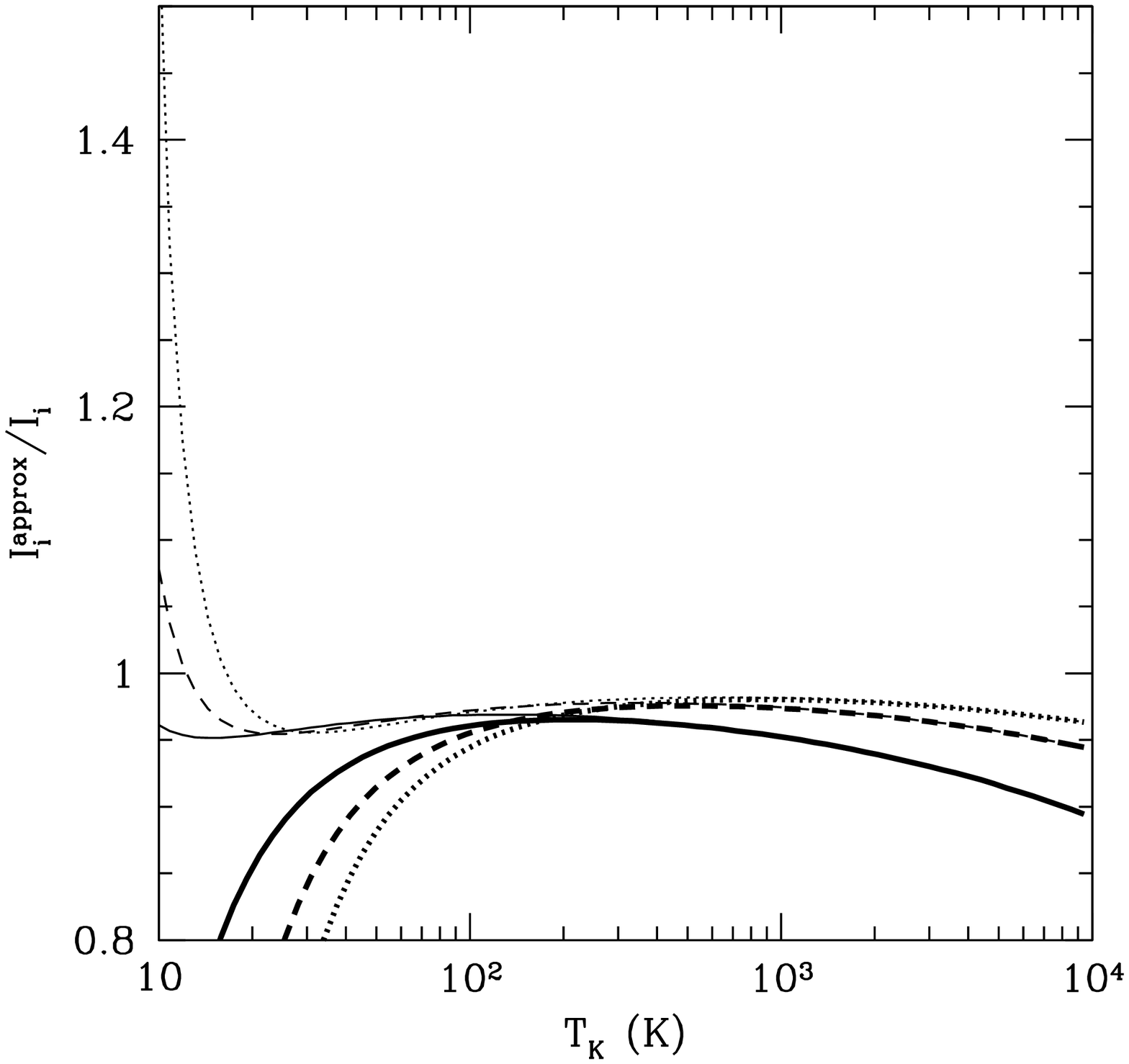}}
\end{center}
\caption{As Fig.~\ref{fig:ictk}, but for injection at line center. In the right panel, the approximate versions include terms up to order $\beta$ (thick curves) and $\beta^2$ (thin curves).}
\label{fig:iitk}
\end{figure*}

\subsection{Injection at Line Center} \label{inj-heat}

For photons injected at the line center, a similar exercise shows that the relevant integral is \citep{chen04}
\begin{equation}
I_i = \int_{-\infty}^0 \deriv x \, \delta_J(x) - \int_0^\infty \deriv x \, \frac{J(x)}{J_{-\infty}}.
\label{eq:iidefn}
\end{equation}
Again, we work in the wing approximation to gain some intuition.  The second integral can be written in closed form; the first is
\bqa
\int_{-\infty}^0 \deriv x \, \delta_J & = & \frac{\eta}{\sqrt{2}} \, \sqrt{\frac{a}{\gamma}} \int_0^{\infty} \frac{\deriv y}{\sqrt{y}} \, \exp \left( - \frac{\pi \gamma}{6 a} y^3 - 2 \eta y \right) \nonumber \\ 
& & \times  {\rm erfc} \left( \sqrt{\frac{\pi \gamma}{2 a} y^3} \right).
\label{eq:iinj-first}
\eqa
Unfortunately, the complementary error function prevents a closed form solution.  However, note that the exponential term implies that the integral is dominated by the region where the argument of the error function is small. Expanding it to lowest order, we then obtain a power series solution in $\beta$:
\begin{equation}
I_i \approx \left( \frac{a}{\gamma} \right)^{1/3}  \sum_{i=0}^\infty A_i \beta^i.
\label{eq:iiapprox}
\end{equation}
The first few terms have $(A_0,\,A_1,\,A_2)=(-0.6979,\,2.5424,\,-2.5645)$.  Retaining only the zeroth-order term, the scaling with $\gamma$ and $T_K$ is again close to that proposed by \citet{chuzhoy06-heat} at $T_K \ga 100 \kel$.

\begin{figure}
\begin{center}
\resizebox{8cm}{!}{\includegraphics{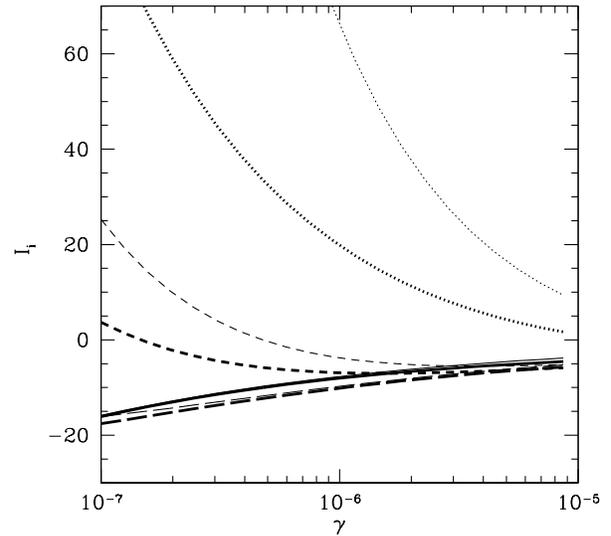}}\\%
\end{center}
\caption{As Fig.~\ref{fig:icgam}, but for injection at line center.  Here the approximate version include terms up to order $\beta$.}
\label{fig:iigam}
\end{figure}

Figures~\ref{fig:iitk} and \ref{fig:iigam} show $I_i$ for the same parameters as in Figures~\ref{fig:ictk} and \ref{fig:icgam}; again we compare the approximate form with the exact solution (including the full Voigt profile).  In this case the dependence on both $T_K$ and $\gamma$ is considerably more complicated.  Most interestingly, injected photons can both heat the gas (when $T_K \la 10 \kel$) and cool it.  Physically, cooling can occur because more photons scatter on the red than the blue side of the line; in such events, the re-emitted photon generally has a higher energy in the IGM frame and so removes heat from the gas.  In the high-temperature regime ($T_K \ga 100 \kel$), the cooling rate falls slightly when $\gamma$ decreases and when $T_K$ increases.  At small temperatures, the exchange switches to heating because the feature is so broad compared to the $\Delta x \sim 1$ frequency change per scattering.

We also show the approximate form (eq.~\ref{eq:iiapprox}) in these panels (note that we must include $\beta^0$ and $\beta^1$ terms).  It is substantially less accurate at first-order in $\beta$, only approaching the exact solution at $T_K \ga 200 \kel$; shortly thereafter, the Voigt profile becomes significant.  However, the thin curves in the right panel of Figure~\ref{fig:iitk} show that carrying the series expansion to $\beta^2$ is quite accurate throughout the range $T_K \ga 10 \kel$.  In the injected case, the wing approximation is less useful because no analytic solution exits.  Thus we recommend numerical integration of equation~(\ref{eq:iidefn}) when high accuracy is required (especially at small temperatures).

\section{Scattering of Ly$n$ Photons}
\label{lyn}

Consider a photon that redshifts into a Ly$n$ line (with frequency $\nu_n$) at redshift $z_r$; its frequency at redshift $z$ is therefore $\nu_z=\nu_n[(1+z)/(1+z_r)]$.  The accumulated optical depth it has traversed by that point is
\begin{equation}
\tau(z) = \int_z^\infty \deriv z \, \frac{\sigma(\nu_z) n_{\rm HI}(z) c}{(1+z) H(z)},
\label{eq:tauz}
\end{equation}
where $\sigma$ refers to the line of interest.  The \citet{gunn65} optical depth is of course the \emph{total} optical depth experienced by such a photon, or $\tau(z \ll z_r)$.  

We are interested in determining the surface at which such a photon will \emph{first} scatter.  For the extremely optically thick Lyman-series lines of the IGM, this first scattering occurs far in the wings of the line, so we can set $\phi(x) \approx a/(\pi x^2)$.  Further assuming the high-redshift limit [$H(z) \propto (1+z)^{3/2}$] and letting $\Delta z \equiv z - z_r \ll z_r$, we have 
\begin{equation}
\tau(z) \approx \frac{n_0^c \chi_{n}}{H_0 \sqrt{\Omega_0}} \, \frac{a}{\pi} \, \frac{(1+z_r)^{5/2}}{\Delta z \, \nu_n^2},
\label{eq:tz2}
\end{equation}
where $\chi_n$ is evaluated for the line of interest.  Re-expressing $\Delta z$ in terms of $x$, we find
\begin{equation}
x(\tau) \approx \frac{1650}{\tau T_K^{1/2}} \, \left( \frac{\nu_\beta}{\nu_n} \right)^4 \, \left( \frac{A_n}{A_\beta} \right) \, \left( \frac{\Gamma_n}{\Gamma_\beta} \right) \left( \frac{1+z_r}{20} \right)^{3/2},
\label{eq:xfirst}
\end{equation}
where we have normalized $\nu$, the spontaneous emission coefficients $A_{n}$, and the inverse lifetimes $\Gamma_n$ to the values appropriate for Ly$\beta$.  Higher Lyman-series photons have significantly longer lifetimes and hence scatter nearer line center; for example, Ly$\epsilon$ photons have a coefficient $\approx 18$.

By setting $\tau=1$ in equation~(\ref{eq:xfirst}), we see immediately that the first scattering occurs well blueward of resonance; we will denote this location $x_1$.  \lya photons cannot be destroyed during scattering (except, of course, in the exceedingly unlikely event that a collision occurs while the atom is excited), so this first scattering limit has little physical interest.  However, higher Lyman-series photons can be destroyed, because the excited state can cascade to an intermediate level.  The destruction probabilities per scattering are compiled by \citet{hirata05} and \citet{pritchard05}; they are $\sim 10\%$ for \lyb and $\sim 20\%$ for higher-level transitions.  Thus, each such photon scatters only a few times before vanishing; so long as they remain in the wings, the $k$th scattering will occur at $x_k \approx x_1/k$.

Because the photons are far out on the blue wing during each of these scattering events, they will deposit some fraction of their energy in the gas, heating it slightly.  Our next goal is to calculate the net energy exchange with the IGM as these photons scatter and eventually disappear.  We begin with the redistribution function $R_{\rm II}(x,x')$, which gives the probability that a photon absorbed at frequency $x'$ is re-emitted at frequency $x$, assuming coherent scattering in the rest frame of the absorbing atom \citep{henyey41, hummer62}, thus ignoring recoil:
\bqa
R_{\rm II}(x,x') & = & \frac{1}{\pi^{3/2}} \int_{|\olx-\ulx|/2}^{\infty} \deriv u \, e^{-u^2} \left[ \tan^{-1} \left( \frac{\ulx+u}{a} \right) \right. \nonumber \\ 
& & \left. - \tan^{-1} \left( \frac{\olx-u}{a} \right) \right],
\label{eq:rtwo}
\eqa
where $\olx={\rm max}(x,x')$ and $\ulx={\rm min}(x,x')$.  This is normalized so that 
\begin{equation}
\phi(x') = \int_{-\infty}^{\infty} \deriv x \, R_{\rm II}(x,x').
\label{eq:rtwonorm}
\end{equation}

In our case ($x' \gg 0$), the redistribution function is sharply peaked around $x'$.  We thus write $x = x' + \Delta x$ and expand the inverse tangents to third order  in $(u,\Delta x)$ about $x'$ (note that, although $u$ can be arbitrarily large, the exponential guarantees that only small values contribute to the integral).  Equation~(\ref{eq:rtwo}) then becomes
\bqa
R_{\rm II}(x,x') & \approx & \frac{a}{\pi^{3/2} x'^2} \int_{|\Delta x|/2}^{\infty} \deriv u \, e^{-u^2} \left[ (2u - \Delta x) \right. \nonumber \\ 
& & \left. + \frac{(\Delta x)^2 - 2 u \Delta x}{x'} \right], \qquad x>x',
\label{eq:rtwo-approx}
\eqa
with a similar expression when $x<x'$.  We are interested in the  mean energy loss in each scattering,
\begin{eqnarray}
\VEV{\Delta x|x'} & = & \phi^{-1}(x') \int_{-\infty}^{\infty} \deriv x \,  \Delta x \, R_{\rm II}(x,x'), \\
& = & \frac{-2}{\pi^{1/2} x'} \int_0^\infty \deriv \Delta x \, \Delta x^2  \nonumber \\ 
& & \times \int_{\Delta x/2}^\infty \deriv u \, e^{-u^2} (2u - \Delta x),
\label{eq:dxmean}
\end{eqnarray}
where we have used the normalization of $R_{\rm II}$ and substituted our expansion for the redistribution function in the second equality.  The integrals are elementary and are most easily performed by switching the order of integration; the simple result is
\begin{equation}
\VEV{\Delta x|x'} = -1/x'.
\label{eq:dx}
\end{equation}
This is identical to the second term in equation~(3) of \citet{chuzhoy06} in the appropriate limit.

As expected, photons tend to lose energy to the gas, but only slowly.  Physically, we have assumed that the scattering is coherent in the rest frame of the atom.  Thus, in the IGM frame, the gas tends to gain energy if the scattering atom travels away from the initial photon and to lose energy if the atom travels toward it.  When scattering occurs blueward of resonance, the former have a slightly higher cross section for absorption because of the small blueshift imparted to them by their thermal velocity; thus the net effect is energy transfer to the gas.  However, far out on the wings of the line, the difference in cross sections from this displacement is small, and the heating is weak.

In contrast, consider scattering with zero natural width (or in other words where the total optical depth is small, so that the initial scattering occurs in the Doppler core).  In that case, $R_{\rm II}$ simplifies to $R_{\rm I}(x,x')={\rm erfc}(\overline{|x|})/2$ (again assuming isotropic scattering in the rest frame), where $\overline{|x|}={\rm max}(|x|,|x'|)$ \citep{unno52}.  This has a flat core between $(-x',x')$, so the typical energy lost in the initial scattering is $\sim x'$.  In this case, photons on the blue side are \emph{only} scattered by atoms moving away from them (so that their frequency lines up with the resonance), and in the lab frame the re-emitted photon typically shifts by a full Doppler width.  Thus, in the few-scatterings limit, heating will be most efficient inside the line core.  This is the case considered (for deuterium) by \citet{chuzhoy06-heat}.

Returning to the Ly$n$ lines, where all the interactions occur in the wings, the net frequency shift in $s$ scattering events is
\begin{eqnarray}
\Delta x_{\rm tot} & \approx & \frac{s(1+s)}{2 x_1},
\label{eq:dxtot} \\
& \sim &  0.033 T_K^{1/2} \left[ \frac{s_n(1+s_n)}{110} \right] \left( \frac{\nu_n}{\nu_\beta} \right)^{4} \left( \frac{A_\beta}{A_n} \right) \nonumber \\ 
& & \times \left( \frac{\Gamma_\beta}{\Gamma_n} \right) \left( \frac{1+z_r}{20} \right)^{-3/2}.
\label{eq:heat-photon}
\end{eqnarray}
Again, we have normalized to the values appropriate for \lyb photons.  Of course, we have assumed that the scatterings occur in the wings of the line.  If the accumulated drift carries the photon toward line center, subsequent scattering will occur symmetrically and heating will be negligible.  Thus we must have $\Delta x_{\rm tot} \le x_1$.  This is marginally true for Ly$\epsilon$, which has $s_\epsilon=5$ and a coefficient $0.77T_K^{1/2}$ when appropriate values are inserted into equation~(\ref{eq:heat-photon}).   

It is useful to compare this drift to that due to recoil itself, which we ignored by using $R_{\rm II}$.  This has $\Delta x_{\rm recoil} = \eta$ per scattering.  Thus
\bqa
\frac{\Delta x_{\rm tot}}{\Delta x_{\rm recoil}} & \sim & 0.11 T_K \left( \frac{1 + s_n}{10} \right) \left( \frac{\nu_n}{\nu_\beta} \right)^{4} \left( \frac{A_\beta}{A_n} \right) \left( \frac{\Gamma_\beta}{\Gamma_n} \right) \nonumber \\ 
& & \times \left( \frac{1+z_r}{20} \right)^{-3/2};
\label{eq:drift-comp}
\eqa
the coefficient is $2.35$ for Ly$\epsilon$ photons.  Thus, at reasonably large temperatures, the frequency drift from repeated scattering overwhelms recoil.  However, unless the IGM is already warm, it never dominates by a large factor.  \citet{pritchard05} showed that recoil provides a negligibly small contribution at all temperatures.  Thus, in practice heating from Ly$n$ scattering is never significant:  only if $T_K \gg T_\gamma$ could it possibly matter, but in that case other, much stronger, heating agents must already be present.

As we have shown, $\Delta x$ is largest ($\sim 1$) when scattering occurs near the line core.  One example, considered by \citet{chuzhoy06-heat}, is the deuterium \lyb resonance, for which the optical depth is of order unity.  Then the energy transfer is much more efficient.  This situation has different temperature dependence.  In either case, $\Delta T \propto h \Delta \nu_D \VEV{\Delta x}$ per scattering.  When absorption is in the line center $\VEV{\Delta x} \sim 1$; in the wings, we have seen that natural broadening controls the cross section and $\VEV{\Delta x} \propto 1/x_1 \propto \Delta \nu_D$.  So $\Delta T_{\rm core} \propto T_K^{1/2}$ and $\Delta T_{\rm wing} \propto T_K$.  Deuterium \lyb turns out to be the most important transition (aside from hydrogen Ly$\alpha$) in heat exchange.  Of course, this energy is injected into the deuterium, rather than the hydrogen, to which it must be transferred by collisions.  According to \citet{chuzhoy06}, this is relatively inefficient, so \lya heating still dominates by a large factor.  As a result, the deuterium temperature may become quite large ($T \sim 10^4 \kel$), where the wing approximation breaks down and the full Voigt profile must be used.

\section{Discussion}
\label{disc}

We have examined both analytic and numeric solutions for the radiation field near the \lya resonance and used them to compute the total scattering rate and the IGM heating (or cooling) rate.  We showed that the approximate analytic solution of \citet{grachev89} and \citet{chuzhoy06}, in which scattering in the wings dominates, is accurate so long as $T_K \la 1000 \kel$.  At higher temperatures, thermal broadening becomes important.  Fortunately, the scattering correction $S_\alpha \rightarrow 1$ at large temperatures.  So the approximate fit presented by \citet{chuzhoy06} -- our equation~(\ref{eq:csfit}) -- turns out to be reasonably accurate (to several percent) whenever $T_K \ga 1 \kel$.  For higher accuracy, equation~(\ref{eq:s-wing}) can be used.

We then used this analytic solution to examine the heating (or cooling) from the scattering near line center.  For the case of photons that redshift toward the resonance, we obtained a fully analytic solution (in terms of Airy functions) under the approximation that all scattering occurs in the wings.  The arguments of the Airy functions are typically small, so a power series expansion is illuminating; it shows that the heating rate per atom and per Hubble time is proportional to $T_K^{-1/3} (1+z)^{-2}$; this is an excellent approximation at $T_K \ga 10 \kel$.  In the case of photons injected at line center, we obtained a power series solution that converges reasonably rapidly.  The lowest order term is reasonably accurate for $T_K \ga 100 \kel$, and in this regime the \emph{cooling} rate per atom and per Hubble time is proportional to $T_K^{1/3} (1+z)^{-5/2}$.  Photons injected in this way only heat the gas when $T_K \la 10 \kel$.

We have neglected drift and diffusivity caused by spin exchange during \lya scattering.  This is relatively easy to incorporate into the approximate solution \citep{chuzhoy06}.  However, the spin temperature actually depends on the spectrum around \lyans, so an iterative process must be employed when $T_K$ is small \citep{hirata05}.  The steps are outlined in \citet{furl06-review}.

As a final thought, it is useful to estimate the heating rate from \lya scattering to gauge its importance relative to other processes.  For simplicity we will consider continuous injection (i.e., photons redshifting into the \lya resonance).  Inserting our lowest order approximation for $I_c$ (from eq.~\ref{eq:icapprox}) into equation~(\ref{eq:heatrate}), we find
\begin{equation}
\frac{2}{3} \frac{\epsilon_\alpha}{H n_{\rm HI} k_B T_K} \approx \frac{0.80}{T_K^{4/3}} \, \frac{x_\alpha}{S_\alpha} \left( \frac{10}{1+z} \right),
\label{eq:lyaheat}
\end{equation}
where the left hand side is the fractional temperature change per Hubble time.  On the right hand side, we have rewritten $J(\nu)$ in terms of the 21 cm coupling efficiency $x_\alpha$ (see eq.~\ref{eq:xalpha}).  The 21 cm spin temperature departs from the CMB temperature when $x_\alpha \sim 1$, so this is a convenient gauge for the background fluxes at which heating is relevant.  Clearly, \lya heating is negligible unless the initial temperature is also small.  (Note that this approximation for $I_c$ \emph{overestimates} the heating at low temperatures, so the actual heating is even smaller than predicted by eq.~\ref{eq:lyaheat}.)  Because, even without any heating, $T_K = 2.5 \kel$ at $z=10$ \citep{seager99}, \lya scattering is unlikely to be significant in this context.  Other processes, especially X-ray heating, are probably much more important \citep{oh01, glover03, furl06-glob}.  

We thank G. Rybicki and M. Furlanetto for helpful discussions.


\begin{thebibliography}{}

\bibitem[\protect\citeauthoryear{{Basko}}{{Basko}}{1981}]{basko81}
{Basko} M.~M.,  1981, Astrophysics, 17, 69

\bibitem[\protect\citeauthoryear{{Chen} \& {Miralda-Escud{\' e}}}{{Chen} \&
  {Miralda-Escud{\' e}}}{2004}]{chen04}
{Chen} X.,  {Miralda-Escud{\' e}} J.,  2004, \apj, 602, 1

\bibitem[\protect\citeauthoryear{{Chugai}}{{Chugai}}{1980}]{chugai80}
{Chugai} N.~N.,  1980, Soviet Astronomy Letters, 6, 91

\bibitem[\protect\citeauthoryear{{Chugai}}{{Chugai}}{1987}]{chugai87}
{Chugai} N.~N.,  1987, Astrofizika, 26, 89

\bibitem[\protect\citeauthoryear{{Chuzhoy} \& {Shapiro}}{{Chuzhoy} \&
  {Shapiro}}{2005}]{chuzhoy06}
{Chuzhoy} L.,  {Shapiro} P.~R.,  2005, submitted to \apj \ (astro-ph/0512206)

\bibitem[\protect\citeauthoryear{{Chuzhoy} \& {Shapiro}}{{Chuzhoy} \&
  {Shapiro}}{2006}]{chuzhoy06-heat}
{Chuzhoy} L.,  {Shapiro} P.~R.,  2006, submitted to \apj \ (astro-ph/0604483)

\bibitem[\protect\citeauthoryear{{Field}}{{Field}}{1958}]{field58}
{Field} G.~B.,  1958, Proc. I.R.E., 46, 240

\bibitem[\protect\citeauthoryear{{Furlanetto}}{{Furlanetto}}{2006}]{furl06-glo%
b}
{Furlanetto} S.,  2006, submitted to \mnras \ (astro-ph/0604040)

\bibitem[\protect\citeauthoryear{{Furlanetto}, {Oh} \& {Briggs}}{{Furlanetto}
  et~al.}{2006}]{furl06-review}
{Furlanetto} S.~R.,  {Oh} S.~P.,    {Briggs} F.~H.,  2006, Physics Reports, in
  preparation

\bibitem[\protect\citeauthoryear{{Glover} \& {Brand}}{{Glover} \&
  {Brand}}{2003}]{glover03}
{Glover} S.~C.~O.,  {Brand} P.~W.~J.~L.,  2003, \mnras, 340, 210

\bibitem[\protect\citeauthoryear{{Grachev}}{{Grachev}}{1989}]{grachev89}
{Grachev} S.~I.,  1989, Astrofizika, 30, 347

\bibitem[\protect\citeauthoryear{{Gunn} \& {Peterson}}{{Gunn} \&
  {Peterson}}{1965}]{gunn65}
{Gunn} J.~E.,  {Peterson} B.~A.,  1965, \apj, 142, 1633

\bibitem[\protect\citeauthoryear{{Henyey}}{{Henyey}}{1941}]{henyey41}
{Henyey} L.~G.,  1941, Proc. Nat. Acad. Sci., 26, 50

\bibitem[\protect\citeauthoryear{{Hirata}}{{Hirata}}{2006}]{hirata05}
{Hirata} C.~M.,  2006, \mnras, 367, 259

\bibitem[\protect\citeauthoryear{{Hummer}}{{Hummer}}{1962}]{hummer62}
{Hummer} D.~G.,  1962, \mnras, 125, 21

\bibitem[\protect\citeauthoryear{{Hummer} \& {Rybicki}}{{Hummer} \&
  {Rybicki}}{1992}]{hummer92}
{Hummer} D.~G.,  {Rybicki} G.~B.,  1992, \apj, 387, 248

\bibitem[\protect\citeauthoryear{{Madau}, {Meiksin} \& {Rees}}{{Madau}
  et~al.}{1997}]{madau97}
{Madau} P.,  {Meiksin} A.,    {Rees} M.~J.,  1997, \apj, 475, 429

\bibitem[\protect\citeauthoryear{{Meiksin}}{{Meiksin}}{2006}]{meiksin06}
{Meiksin} A.,  2006, submitted to \mnras \ (astro-ph/0603855)

\bibitem[\protect\citeauthoryear{{Oh}}{{Oh}}{2001}]{oh01}
{Oh} S.~P.,  2001, \apj, 553, 499

\bibitem[\protect\citeauthoryear{{Pritchard} \& {Furlanetto}}{{Pritchard} \&
  {Furlanetto}}{2006}]{pritchard05}
{Pritchard} J.~R.,  {Furlanetto} S.~R.,  2006, \mnras, 367, 1057

\bibitem[\protect\citeauthoryear{{Rybicki}}{{Rybicki}}{2006}]{rybicki06}
{Rybicki} G.~B.,  2006, submitted to \apj \ (astro-ph/0603047)

\bibitem[\protect\citeauthoryear{{Rybicki} \& {dell'Antonio}}{{Rybicki} \&
  {dell'Antonio}}{1994}]{rybicki94}
{Rybicki} G.~B.,  {dell'Antonio} I.~P.,  1994, \apj, 427, 603

\bibitem[\protect\citeauthoryear{{Seager}, {Sasselov} \& {Scott}}{{Seager}
  et~al.}{1999}]{seager99}
{Seager} S.,  {Sasselov} D.~D.,    {Scott} D.,  1999, \apjl, 523, L1

\bibitem[\protect\citeauthoryear{{Spergel} et~al.,}{{Spergel}
  et~al.}{2006}]{spergel06}
{Spergel} D.~N.,  et~al., 2006, submitted to \apj \ (astro-ph/0603449)

\bibitem[\protect\citeauthoryear{{Unno}}{{Unno}}{1952}]{unno52}
{Unno} W.,  1952, \pasj, 3, 158

\bibitem[\protect\citeauthoryear{{Wouthuysen}}{{Wouthuysen}}{1952}]{wouthuysen%
52}
{Wouthuysen} S.~A.,  1952, \aj, 57, 31

\end{thebibliography}

\end{document}